\documentclass{article}
\usepackage{spconf,amsmath,graphicx,hyperref}
\usepackage{amssymb}
\usepackage{amsfonts}
\usepackage{booktabs}  
\usepackage{enumitem}
\usepackage{array}
\usepackage{tabularx}
\usepackage{cite}
\usepackage{amsfonts}
\usepackage{siunitx}
\usepackage{caption}
\makeatletter
\renewcommand\normalsize{%
   \@setfontsize\normalsize{9pt}{10.5pt}} 
\makeatother
\let\OLDthebibliography\thebibliography
\renewcommand\thebibliography[1]{
  \OLDthebibliography{#1}
  \setlength{\parskip}{0.5pt}
  \setlength{\itemsep}{1pt plus 0.3ex}
}
\usepackage{flushend}

\def\H{{\mathsf H}}

\title{VM-UNSSOR: Unsupervised Neural Speech Separation Enhanced by Higher-SNR Virtual Microphone Arrays}
\name{Shulin He and Zhong-Qiu Wang$^{*}$
\thanks{$^{*}$Corresponding author: Zhong-Qiu Wang.}
}
\address{
Southern University of Science and Technology, Shenzhen, China\\
{\small\texttt{\{he.shulin96, wang.zhongqiu41\}@gmail.com}}
}
\begin{document}
%
\maketitle
\begin{abstract}
Blind speech separation (BSS) aims to recover multiple speech sources from multi-channel, multi-speaker mixtures under unknown array geometry and room impulse responses.
In unsupervised setup where clean target speech is not available for model training, UNSSOR proposes a mixture consistency (MC) loss for training deep neural networks (DNN) on over-determined training mixtures to realize unsupervised speech separation.
However, when the number of microphones of the training mixtures decreases, the MC constraint weakens and the separation performance falls dramatically.
To address this, we propose VM-UNSSOR, augmenting the observed training mixture signals recorded by a limited number of microphones with several higher-SNR virtual-microphone (VM) signals, which are obtained by applying linear spatial demixers (such as IVA and spatial clustering) to the observed training mixtures.
As linear projections of the observed mixtures, the virtual-microphone signals can typically increase the SNR of each source and can be leveraged to compute extra MC losses to improve UNSSOR and address the frequency permutation problem in UNSSOR.
On the SMS-WSJ dataset, in the over-determined six-microphone, two-speaker separation setup, VM-UNSSOR reaches $17.1$ dB SI-SDR, while UNSSOR only obtains $14.7$ dB; and in the determined two-microphone, two-speaker case, UNSSOR collapses to $-2.7$ dB SI-SDR, while VM-UNSSOR achieves $10.7$ dB.
\end{abstract}
\section{Introduction}

The cocktail party problem \cite{cherry1953some,Wang2006Book,McDermott2009} arises when several people speak at the same time in the same environment, wherein the microphones inevitably record a mixture of all the concurrent speech.
This problem is widely encountered, especially in applications such as smart speakers, smart cockpit (in electric vehicles), and wearable devices such as smart glasses \cite{cherry1953some,haykin2005cocktail}.
The goal of speech separation is to separate the mixture and recover each individual speaker signal so that downstream speech understanding applications such as automatic speech recognition, speaker identification, and hearing assistance can work robustly.
Supervised speech separation based on deep neural networks (DNN) obtains impressive performance nowadays when the training and test conditions are matched with each other \cite{wang2018supervised,hershey2016deep,kolbaek2017multitalker,vzmolikova2019speakerbeam,wang2019deep,luo2019conv,liu2019divide,wang2018combining,luo2018speaker, 10389733}, yet the performance often drops dramatically in unseen acoustic environments.
Meanwhile, collecting paired clean source signals and mixtures for every scenario is costly and in many cases impractical.
These challenges motivated recent research on unsupervised speech separation (USS), which directly train DNNs on unlabeled mixtures recorded in the target environment via unsupervised learning \cite{wang2023unssor,wisdom2020unsupervised,saijo2023self,wang2017gender,drude19_interspeech,sivaraman2022adapting,seetharaman2019bootstrapping}.

UNSSOR \cite{wang2023unssor}, a recent algorithm in this line of research, proposes a so-called mixture consistency (MC) loss afforded by over-determined training mixtures to realize USS.
The insight is that the source estimates, after being properly linearly filtered, should be able to reconstruct the observed mixture at each microphone.
In other words, the mixture signal at each microphone can be leveraged as a constraint to regularize the source estimates, thereby realizing separation.
In detail, during training, a DNN is trained to produce an estimate for each speaker, and, for the mixture signal at each microphone, the estimates of the speakers are linear-filtered and summed up to minimize the distance between the summated signal and the mixture signal (i.e., MC loss).
This procedure yields supervision without requiring clean reference signals to penalize the DNN estimates \cite{wang2023unssor}.
Clearly, the supervision would become stronger when microphones outnumber speakers, because each additional microphone introduces one more MC constraint that could benefit training.
For training mixtures recorded by microphone arrays with a limited number of microphones,
the constraints are weaker and separation quality would degrade dramatically.

To address these limitations, we propose to introduce virtual microphones that increase the effective microphone count without using additional hardware (i.e., physical microphones).
We denote $\mathcal{V}$ as the set of virtual microphones, derived by applying linear spatial demixers such as independent vector analysis (IVA) \cite{sawada2019review,kim2006independent,hiroe2006solution,boeddeker2021comparison} or spatial clustering (SC) \cite{rickard2007duet,vu2010blind,sawada2010underdetermined,ito2016complex} to the observed mixture signals.
Because each virtual microphone in $\mathcal{V}$ is a linear projection of the observed mixture signals, it follows the same acoustic mixing model and can be used directly to compute additional MC loss for unsupervised training.
On the other hand, as the linear spatial demixers are usually effective to some extent, $\mathcal{V}$ often exhibit higher SNR for the sources.
Such higher-SNR signals could act as a pseudo-teacher to help separate the observed lower-SNR mixture signals, and, in addition, could help solve the frequency permutation problem \cite{sawada2019review}, a unique issue that needs to be addressed in USS.
We leverage the virtual-microphone signals to compute additional MC losses to penalize the DNN estimates, instead of having the estimates to directly fit the virtual-microphone signals, considering that doing so would limit the DNN's separation capability by that of the linear spatial demixers.

Linear spatial demixers such as IVA \cite{sawada2019review,kim2006independent,hiroe2006solution,boeddeker2021comparison} often back-project separated signals to the reference microphone.
This would skew the overall MC loss towards that microphone and degrades training stability.
We mitigate this issue by back-projecting the separated estimates to every physical microphone. That is, for each physical microphone, the number of virtual microphones we create equals the number of sources.
We then apply a re-weighted MC loss that balances the contributions of physical and virtual microphones.
During training, we compute the MC loss on all microphones, physical and virtual.
This increases the number of constraints.
The separator takes as input the concatenated physical and virtual microphones.

We name the proposed system \textit{VM-UNSSOR}.
By injecting virtual microphones, determined mixtures become pseudo over-determined during training, and over-determined mixtures gain extra constraints.
This way, we can keep the training of the system label-free and not requiring additional hardware microphones.
The contributions of this work can be summarized as follows:
\begin{itemize}[leftmargin=*,noitemsep,topsep=0pt]
\item We extend UNSSOR by introducing virtual microphones whose signals offer higher-SNR cues that strengthen the MC constraint.
\item We show that a simple physical–virtual re-weighted MC loss enables unsupervised training on determined trainining mixtures by creating pseudo over-determined constraints.
\item On the SMS-WSJ dataset \cite{drude2019sms}, VM-UNSSOR achieves $17.1$ dB SI-SDR in the six-microphone, two-speaker setup, compared with $14.7$ dB obtained by UNSSOR; and in the determined two-microphone, two-speaker setup, UNSSOR fails to train ($-2.7$ dB SI-SDR) while VM-UNSSOR reaches $10.7$ dB.
\end{itemize}

\section{Background}
\label{sec:background}

\subsection{Notations and Speaker–Image Physical Model}

We operate in the short-time Fourier transform (STFT) domain.
Let $p\in\{1,\dots,P\}$ indexes $P$ microphones, $c\in\{1,\dots,C\}$ indexes $C$ speakers, $t$ indexes $T$ frames, and $f$ indexes $F$ frequency bins.
UNSSOR models each microphone signal as a sum of speaker images:
\begin{equation}
\label{eq:mixing}
Y_p(t,f) = \sum\nolimits_{c=1}^{C} X_p(c,t,f) + \varepsilon_p(t,f),
\end{equation}
where $X_p(c)$ is the reverberant image of speaker $c$ at microphone $p$, and $\varepsilon_p$ absorbs residual noises and modeling error.
Without loss of generality, microphone~$1$ is designated as the reference microphone.
The goal is to estimate the speaker images $\{X_1(c)\}_{c=1}^{C}$ at the reference microphone in an unsupervised manner while preserving their reverberation.

Let $P$ denote the number of physical microphones, and we define the size of the full microphone index set as $\mathcal{P}=\{1,\dots,P\}$.
Let $\mathcal{R}\subseteq\mathcal{P}$ denote the subset of physical microphones whose mixture signals are actually used as the input to the separator, and we denote its cardinality as $P_r$ ($=|\mathcal{R}|$).
Virtual microphones constructed by spatial demixers are collected in the set $\mathcal{V}$ with size $Q$ ($=|\mathcal{V}|$).
We use $\mathcal{U} = \mathcal{R} \cup \mathcal{V}$ as the combined input set, so $|\mathcal{U}| = P_r + Q$.
In particular, we do not overload the notation $P$ to mean $|\mathcal{U}|$ or $P_r$, and $P$ always refers to the total number of physical microphones.

\subsection{Relative RIR Constraint via Short Convolutions}

For a small-aperture microphone array, the images of the same speaker at nearby microphones can be well-approximated by using a short linear filter between the microphones \cite{Gannot2017}.
Let $p\in\{1,\dots,P\}$ index microphones and fix $p=1$ as the reference.
We first define a reference-image temporal context
\begin{equation}
\tilde X_1(c,t,f)=\big[X_1(c,t-A,f),\dots,X_1(c,t+B,f)\big]^{\top}\in\mathbb{C}^{E},\label{temporal_context}
\end{equation}
with $E=A+B+1$. Then, for each non-reference microphone $p$ (where $p\neq 1$) there exists a relative RIR $g_p(c,f)\in\mathbb{C}^{E}$ such that
\begin{equation}
\label{eq:relrir}
X_p(c,t,f)\;\approx\; g_p(c,f)^{\H}\,\tilde X_1(c,t,f),
\end{equation}
where $(\cdot)^{\H}$ computes Hermitian transpose. In other words, at each frequency $f$, $X_p(c,\cdot,f)$ can be modeled as a short convolution of the reference image $X_1(c,\cdot,f)$ and a relative RIR $g_p(c,f)$.


\subsection{UNSSOR Training Mechanism}

A neural separator $g_{\theta}$ is trained to output a complex-valued estimate $\hat Z(c)$ for each speaker $c$.
Given $\hat Z(c)$ and the observed mixture signal $Y_p$, UNSSOR estimates the relative RIR by forward convolutive prediction (FCP) \cite{wang2023unssor,wang2025superm2m,wang2024usdnet,wang2024cross}:
\begin{equation}
\label{eq:fcp}
\hat g_p(c,f) = \arg\min_{g_p(c,f)}\sum_{t}\Big|Y_p(t,f)-g_p(c,f)^{\H}\tilde{\hat Z}(c,t,f)\Big|^{2},
\end{equation}
which has a closed-form solution.
Next, the FCP-estimated speaker image at microphone $p$ is computed via
\begin{equation}
\hat X^{\text{FCP}}_{p}(c,t,f)=\hat g_p(c,f)^{\H}\tilde{\hat Z}(c,t,f).\label{FCP_results}
\end{equation}
UNSSOR \cite{wang2023unssor} defines a label-free reconstruction loss named MC loss, which enforces consistency between the mixture signal and the summation of the FCP-estimated speaker images at each microphone. That is, 
\begin{align}
\mathcal{L}_{\mathrm{MC}}=\sum_{p=1}^{P}\mathcal{L}_{\mathrm{MC},p},\label{overall_MC_Loss}
\end{align}
with $\mathcal{L}_{\mathrm{MC},p}$ denoting the MC loss at microphone $p$ and defined as
\begin{align}
\mathcal{L}_{\mathrm{MC},p} = &\sum_{t,f}\Big(
w_r\times\Big|\mathcal{R}\Big(
Y_p(t,f) - \sum\nolimits_{c} \hat X^{\mathrm{FCP}}_{p}(c,t,f)
\Big)\Big| \nonumber \\
& + w_i\times\Big|\mathcal{I}\Big(
Y_p(t,f) - \sum\nolimits_{c} \hat X^{\mathrm{FCP}}_{p}(c,t,f)
\Big)\Big| \nonumber \\
& + w_m\times\Big||Y_p(t,f)| - |\sum\nolimits_{c}
\hat X^{\mathrm{FCP}}_{p}(c,t,f)|\Big|
\Big),\label{eq:mc}
\end{align}
where $|\cdot|$ extracts magnitude, $\mathcal{R}(\cdot)$ and $\mathcal{I}(\cdot)$ respectively extract real and imaginary components, and $(w_r,w_i,w_m)$ are weighting terms controlling the contributions of the losses on the real component, imaginary component, and magnitude.
We implement per-microphone energy normalization following UNSSOR \cite{wang2023unssor}.

Since $\hat g_p(c,f)$ is estimated independently per frequency, cross-frequency permutations can occur.
To address this, UNSSOR includes an intra-source magnitude scattering (ISMS) loss to promote consistent spectral patterns across frequencies \cite{wang2023unssor}:
\begin{equation}\label{eq:isms}
\mathcal{L}_{\text{ISMS}}
=\frac{\sum_t \frac{1}{C} \sum_{c=1}^C \text{var}\Big(\text{log}(|\hat{X}_p^{\text{FCP}}(c,t,\cdot)|)\Big)}{\sum_t \text{var}\Big(\text{log}(|Y_p(t,\cdot)|)\Big)},
\end{equation}
where $\text{var}(\cdot)$ computes the variance of the values in a vector.


\begin{figure}[!t]
    \centering
    \includegraphics[width=0.5\textwidth, keepaspectratio]{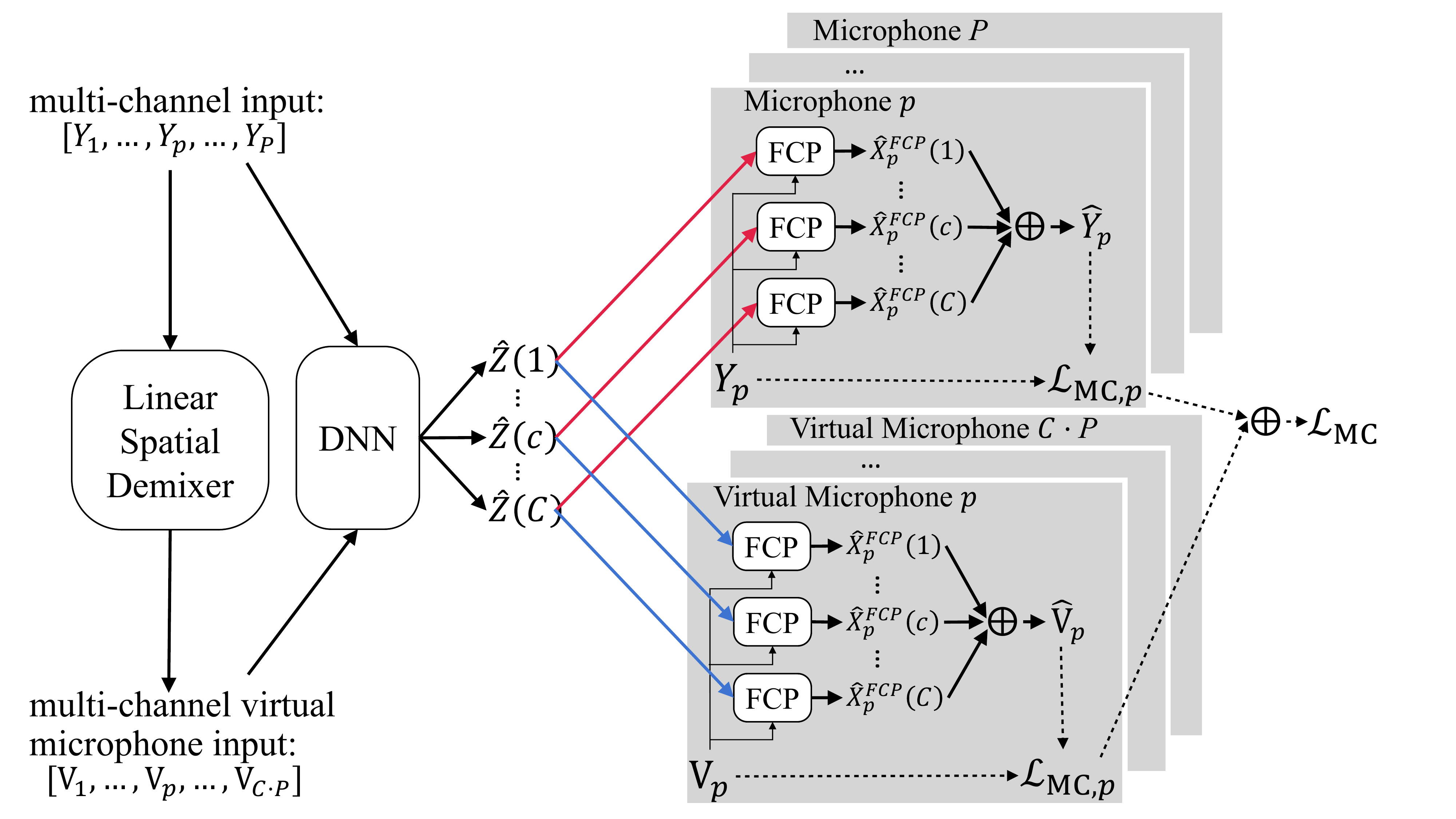}
    \caption{Overview of VM-UNSSOR.
    A linear spatial demixer derives $\mathcal{V}$ via back-projection.
    The separator DNN takes the physical and virtual channels as input, using FCP and MC losses to enforce per-channel consistency.}
    \label{fig:overview}
\end{figure}
 
\section{VM-UNSSOR}\label{sec:method}

Fig.~\ref{fig:overview} shows VM-UNSSOR, where a linear spatial demixer computes higher-SNR VM signals.
We feed the concatenated physical and virtual microphone signals as inputs to the neural separator, and leverage FCP to enforce per-microphone mixture consistency.

\subsection{Virtual Microphones from Linear Spatial Demixers}
\label{sec:vm_synth}
Let $\mathcal{R}=\{1,\dots,P_r\}$ denote the set of physical microphones used as the input to the DNN.
We synthesize virtual microphones in the STFT domain by applying linear spatial demixers to the physical-array mixture signals.
Concretely, we estimate a frequency-wise demixing matrix $W(f)\in\mathbb{C}^{C\times P_r}$ using IVA on the raw mixtures and obtain separated components:
\begin{equation}
\hat S_c(t,f)=w_c(f)^{\H}Y_{\mathcal{R}}(t,f),\, \text{for} \, c=1,\dots,C,
\end{equation}
where $Y_{\mathcal{R}}(t,f)\in\mathbb{C}^{P_r}$ stacks the $P_r$ physical microphones and $w_c(f)$ is the $c$-th row of $W(f)$. We then form a mixing estimate $A(f)\in\mathbb{C}^{P_r\times C}$ (i.e., the pseudo-inverse of $W(f)$) and back-project each separated component to every physical microphone:
\begin{equation}
V_{p,c}(t,f)=A_{p,c}(f)\,\hat S_c(t,f),\,\text{for}\, p=1,\dots,P_r,\; c=1,\dots,C.
\end{equation}
Each $V_{p,c}$ is a linear combination of the physical microphones. Therefore, it is consistent with the same acoustic mixing model as $Y_p$. This construction yields $Q=C\cdot P_r$ virtual microphones.

Let $\mathcal{V}=\{(p,c)\,:\,p\in\mathcal{R},\,c\in\{1,\dots,C\}\}$ denotes the set of virtual microphones, and let $\mathcal{U}=\mathcal{R}\cup\mathcal{V}$ be the augmented observation stack.
We define
\begin{equation}
\label{eq:aug_obs}
O_k(t,f)=
\begin{cases}
Y_k(t,f), & k\in\mathcal{R},\\[2pt]
V_{p,c}(t,f), & k=(p,c)\in\mathcal{V},
\end{cases}
\end{equation}
and feed the physical and virtual (mixture) signals $\{O_k\}_{k\in\mathcal{U}}$ to the separator $g_\theta$. The total number of input signals is $P_u=P_r+Q=P_r\times (1+C)$.

\subsection{Mixture Consistency Loss on Virtual Microphones}
\label{sec:consistency}
Let $\hat Z(c)$ be the estimate for source $c$ produced by the neural separator $g_\theta$ based on the augmented mixture signals. For each microphone $k\in\mathcal{U}$ and frequency $f$, we estimate a microphone-specific relative filter $\hat g_k(c,f)\in\mathbb{C}^{E}$ by FCP:
\begin{equation}
\label{eq:fcp_k}
\begin{aligned}
\hat g_k(c,f) = \operatorname*{arg\,min}_{g_k(c,f)} \sum_{t} \Bigl|O_k(t,f) - g_k(c,f)^{\H}\tilde{\hat Z}(c,t,f) \Bigr|^{2},
\end{aligned}
\end{equation}
where $\tilde{\hat Z}(c,t,f)$ is the length-$E$ temporal context of $\hat Z(c,t,f)$, constructed by following Eq. (\ref{temporal_context}). The corresponding FCP-estimated image at microphone $k$ is
\begin{equation}
\label{eq:fcp_img_k}
\hat X^{\text{FCP}}_{k}(c,t,f)=\hat g_k(c,f)^{\H}\,\tilde{\hat Z}(c,t,f).
\end{equation}
Aggregating constraints across $\mathcal{U}$
strengthens the overall MC constraint, and in addition leads to more stable FCP estimation.

The training loss is defined as
\begin{equation}
\label{eq:loss_vm}
\begin{aligned}
\mathcal{L}_{\text{VM}} = \alpha\times \sum_{k\in\mathcal{R}} \mathcal{L}_{\mathrm{MC},k} + \beta\times \sum_{k\in\mathcal{V}} \mathcal{L}_{\mathrm{MC},k},
\end{aligned}
\end{equation}
where $\alpha$ and $\beta$ are tunable weighting terms balancing the MC losses on the physical and virtual microphones.

\section{Experimental Setup}
\label{sec:exp}

\begin{table*}[t]
\footnotesize
\centering
\caption{Results on SMS-WSJ ($6$-microphone, $2$-speaker setup).
“Input ch.” are microphones fed to the separator.
“VM-loss (ch.)” counts the number of channels used in the loss.
“-” means no VM ($\beta{=}0$).
“VM-input” uses $\mathcal{V}$ as additional inputs to the separator.
$\alpha$/$\beta$ weight physical/virtual microphones.
“ISMS” shows whether the ISMS loss is enabled.
Virtual microphones are formed by IVA (Gaussian).
In this result, $C{=}2$ and $P_r{=}6$, $Q{=}12$ and $P_u{=}18$.
}
\vspace{-0.2cm}
\label{tab:vm_results}
\setlength{\tabcolsep}{4pt}
\footnotesize
\sisetup{table-format=2.2,round-mode=places,round-precision=2,table-number-alignment = center,detect-weight=true,detect-inline-weight=math}
\begin{tabular}{
l
l
S[table-format=2,round-precision=0] 
S[table-format=2,round-precision=0] 
S[table-format=1.1,round-precision=1] 
S[table-format=1.2,round-precision=2] 
c
S[table-format=2.1,round-precision=1] 
S[table-format=2.1,round-precision=1] 
S[table-format=1.2,round-precision=2] 
S[table-format=1.3,round-precision=3] 
S[table-format=1.3,round-precision=3] 
}
\toprule
Row & Systems & {Input ch.} & {VM-loss (ch.)} & $\alpha$ & $\beta$ & ISMS & {SI-SDR(dB)$\uparrow$} & {SDR(dB)$\uparrow$} & {NB-PESQ$\uparrow$} & {STOI$\uparrow$} & {eSTOI$\uparrow$} \\
\midrule
0a & Mixture (unprocessed)              & {-}  & {-}   & {-} & {-}     & {-}          & -0.03 & 0.06 & 1.87 &  0.6030 & 0.7218 \\
\midrule
1a & Demixer-only baseline \cite{boeddeker2021comparison}  & 6  & {-}   & {-} & {-}     & {-}          & 13.39   & 14.75   & 3.08   & 0.8662 & 0.9477      \\
1b & ArrayDPS \cite{xu2025arraydps}  & 6  & {-}   & {-} & {-}     & {-}          &  16.20  & 16.90  & 3.49   & 0.8840 &    {-}   \\
\midrule
2a & UNSSOR \cite{wang2023unssor}                            & 6  & {-}   & 1 & {-}     & \checkmark & 14.68   & 15.54   & 3.42   & 0.8874 & 0.9562 \\
2b & UNSSOR + VM-loss                   & 6  & 18  & 1 & 0.02  & \checkmark & 14.8583   & 15.7468   & 3.4969   & 0.8927 & 0.9577 \\
2c & UNSSOR + VM-loss                   & 6  & 18  & 1 & 0.02  & $\times$   & 15.30   & 16.17   & 3.49   & 0.9022 & 0.9631 \\
\midrule
3a & UNSSOR + VM-input                  & 18 & {-}   & 1 & {-}     & \checkmark & 16.62   & 17.59   & 3.55   & 0.9118 & 0.9662 \\
3b & UNSSOR + VM-input + VM-loss        & 18 & 18  & 1 & 0.02  & \checkmark & 16.67   & 17.68   & 3.57   & 0.9137 & 0.9665 \\
3c & UNSSOR + VM-input + VM-loss        & 8 & 8  & 1 & 0.02  & \checkmark & 15.5428   & 16.4083   & 3.5236   & 0.9055 & 0.9648 \\
3d & VM-UNSSOR                          & 18 & 18  & 1 & 1     & $\times$ &  14.3354  &  15.8969  & 3.3602   & 0.8854 & 0.9539 \\
3e & VM-UNSSOR        & 18 & 18  & 1 & 0.06  & $\times$   & 16.78   & 17.81   & 3.58   & 0.9150 & 0.9671 \\
3f & VM-UNSSOR          & 18 & 18  & 1 & 0.02  & $\times$ & \bfseries 17.05 & \bfseries 18.04 & \bfseries 3.59 & \bfseries 0.9178 & \bfseries 0.9686 \\
\bottomrule
\end{tabular}
\end{table*}

\subsection{Dataset and Evaluation Setup}
\label{sec:exp:setup}

All experiments are based on the two-speaker SMS-WSJ corpus \cite{drude2019sms}.
We use the same training, validation, and test sets, room simulation settings, and STFT configuration as in UNSSOR \cite{wang2023unssor}.
The UNSSOR baseline is trained and evaluated with $6$ physical microphones.
For VM-UNSSOR, we adopt $P_r = 6$ physical microphones and synthesize $Q=C \times P_r$ virtual microphones per mixture by frequency-wise linear spatial demixers with IVA, which yields $Q = 12$ and an augmented observation stack of $P_u = P_r + Q = 18$ signals for the two-speaker case.
We report averaged SDR \cite{Vincent2006a}, SI-SDR \cite{LeRoux2019}, NB-PESQ \cite{Rix2001}, STOI \cite{STOI}, and eSTOI \cite{7539284} scores on the official test set.
For the two-microphone experiments, we use channels $0$ and $3$ from the $6$-microphone recordings.

\subsection{Baseline Systems}
\label{sec:exp:baselines}

UNSSOR \cite{wang2023unssor} trains the separator using a combination of the MC and loss in Eq. (\ref{overall_MC_Loss}) ISMS loss in (\ref{eq:isms}).
It estimates per-frequency FCP filters from the mixtures and separator's outputs.
We adopt its original setup for SMS-WSJ proposed in \cite{wang2023unssor}.

We utilize IVA and spatial clustering (SC) as the demixers.
For IVA, we implement it with a Gaussian source model using the \emph{torchiva} toolkit \cite{torchiva}.
On over-determined arrays, we run a $3$-source IVA and drop the lowest-energy estimate.
On determined arrays we run a $2$-source IVA.
The STFT uses $256$ ms window and $32$ ms hop size.
For SC, we use a public CACGMM implementation with inter-frequency correlation for frequency alignment \cite{Boeddeker2019SpatialClustering}.
On over-determined arrays we estimate three sources and drop the source with the lowest energy, and on determined arrays we estimate two.
The STFT uses $128$ ms window and $16$ ms hop size.
Unless otherwise stated, VM-UNSSOR adopts IVA to form $Q = C \cdot P_r$ virtual microphones, which we found yields stable demixing on SMS-WSJ.

ArrayDPS \cite{xu2025arraydps} is a generative approach based on diffusion posterior sampling for USS.
It combines a diffusion prior to leverage speech priors and a likelihood based on mixture consistency to satisfy signal regularizations enforced by observed mixtures.
It uses IVA results to initialize its sampling process
and reports results on SMS-WSJ under the same metrics used here.
We cite its published SMS-WSJ configurations and scores for comparison in our tables. 

The training procedure of VM-UNSSOR (e.g., learning-rate scheduling, optimizer, gradient clipping, and data augmentation) follows the UNSSOR recipe.
For VM-UNSSOR, in the re-weighted loss described in Eq. (\ref{eq:loss_vm}), we use $\alpha=1.0$ for physical microphones and $\beta=0.02$ for virtual microphones, unless otherwise noted.

\section{Evaluation Results}
\label{sec:exp:main}

Table~\ref{tab:vm_results} reports two-speaker results with six physical microphones.
Rows 0a and 1a report the results of the unprocessed mixtures and the demixer-only IVA output, and row 2a is the UNSSOR baseline. Adding only the virtual-microphone MC loss while keeping the separator input at six physical microphones (rows 2b/2c) increases SI-SDR from $14.7$ to $14.9$/$15.3$ dB and SDR from $15.5$ to $15.7$/$16.2$ dB (with/without using the ISMS loss, respectively).
Feeding the physical and virtual microphone signals as inputs but without VM-loss (row 3a) further increases SI-SDR to $16.6$ dB and SDR to $17.6$ dB.
Combining VM-input and VM-loss with the ISMS loss enabled (row 3b) yields $16.7$ dB SI-SDR and $17.7$ dB SDR.
The best configuration is in row 3f, where ISMS is disabled and $\beta=0.02$, reaching $17.1$ dB SI-SDR and $18.0$ dB SDR.
In comparison, ArrayDPS in row 1b attains $16.2$ dB SI-SDR and $16.9$ dB SDR.
VM-UNSSOR surpasses its performance in the same six-microphone setting.
All virtual microphones in Table~\ref{tab:vm_results} are formed by IVA with a Gaussian source model.

Row 2b vs. 2a shows that adding virtual microphones only to the loss brings gains without changing the inference input.
Row 3a vs. 2a shows that providing virtual microphones as inputs is also helpful.
Row 3b vs. 3a shows a further improvement when VM-loss is enabled after VM-input is already in place.
Overall, virtual microphones enlarge the set of MC losses and can offer higher-SNR observations that the separator can benefit.

With six physical inputs and VM-loss, turning off ISMS (row 2c vs. 2b) gives a small improvement.
With $18$ inputs (physical + virtual) and VM-loss, turning off ISMS (row 3f vs. 3b) improves SI-SDR from $16.7$ to $17.1$ dB and SDR from $17.7$ to $18.0$ dB.
A possible explanation is that virtual microphones can already provide source dominance and alignment cues that could help resolve the frequency permutation problem.
In this case, including the ISMS loss could make the estimated magnitudes too uniform across frequencies, leaving the separator less able to correct demixer artifacts.

In row 3c, IVA on six microphones yields two source estimates that are both back-projected to the reference microphone $1$, which biases the MC loss and hurts performance.
In 3f, the same IVA outputs are back-projected to all the six microphones, preserving MC balance and giving better results.

Setting $\beta=0.02$ (in row 3f) leads to the best performance. At $\beta=0.06$ (in row 3e), both SI-SDR and SDR decline, and at $\beta=1$ (in row 3d), they fall further.
As $\beta$ increases, the loss over-weights the virtual microphones.
Because they inherit demixer imperfections, the separator starts fitting demixing artifacts instead of enforcing mixture consistency on all channels.

\begin{table}[t]
\centering
\caption{Demixing method for virtual microphones on SMS-WSJ ($6$-microphone, $2$-speaker setup). “Demixer-only” means using demixers alone.}
\label{tab:demix_choice}
\vspace{-0.2cm}
\sisetup{table-format=2.2,round-mode=places,round-precision=2,table-number-alignment = center,detect-weight=true,detect-inline-weight=math}
\setlength{\tabcolsep}{4pt}
\footnotesize
\begin{tabular}{
lc
S[table-format=2,round-precision=0] 
S[table-format=2.1,round-precision=1] 
}
\toprule
Systems & Demixer & {Input ch.} & {SI-SDR(dB)$\uparrow$} \\
\midrule

Demixer-only baseline \cite{boeddeker2021comparison} & SC  (6 mics) & {-} & 7.39 \\
Demixer-only baseline \cite{boeddeker2021comparison} & IVA (6 mics) & {-} & 13.39 \\
\addlinespace[2pt]
VM-UNSSOR & SC            & 18 & 16.90 \\
VM-UNSSOR & IVA &  18 & \bfseries 17.05 \\
\bottomrule
\end{tabular}
\end{table}

Table~\ref{tab:demix_choice} compares using IVA and spatial clustering as the linear demixers for forming virtual microphones.
The demixer-only rows measure the demixers by themselves without VM-UNSSOR.
IVA itself obtains $13.4$ dB SI-SDR, which is higher than $7.4$ dB for spatial clustering.
For VM-UNSSOR, using an IVA frontend yields $17.1$ dB SI-SDR, whereas using a spatial clustering frontend yields $16.9$ dB.
These results indicate that VM-UNSSOR is compatible with different demixers and that better demixing quality leads to better separation, since higher-SNR virtual microphones can strengthen mixture consistency and provide clearer source dominance cues for learning.

\begin{table}[t]
\centering
\caption{$2$-microphone, $2$-speaker results on SMS-WSJ. “Demixer-only” means using demixer alone.
``Input ch.=$6$'' means $2$ physical plus $4$ virtual microphones.}
\vspace{-0.2cm}
\label{tab:vm_2mic}
\setlength{\tabcolsep}{4pt}
\footnotesize
\sisetup{table-format=2.2,round-mode=places,round-precision=2,table-number-alignment = center,detect-weight=true,detect-inline-weight=math}
\begin{tabular}{lc
S[table-format=2,round-precision=0] 
S[table-format=2.1,round-precision=1] 
}
\toprule
Systems & Demixer & {Input ch.} & {SI-SDR(dB)$\uparrow$} \\
\midrule
Demixer-only baseline \cite{boeddeker2021comparison} & SC (2 mics) & {-} & 6.24 \\
Demixer-only baseline \cite{boeddeker2021comparison} & IVA (2 mics) & {-} & 9.14 \\
\addlinespace[2pt]
UNSSOR & {-} & 2 & -2.70 \\
VM-UNSSOR & SC & 6 & -0.75 \\
VM-UNSSOR & IVA & 6 & \bfseries 10.72 \\
\bottomrule
\end{tabular}
\end{table}

Table~\ref{tab:vm_2mic} reports the determined two-microphone, two-speaker setting.
UNSSOR fails to train and obtains $-2.7$ dB SI-SDR.
The demixer-only rows show IVA at $9.1$ dB and spatial clustering at $6.2$ dB.
With VM-UNSSOR, IVA-based virtual microphones achieve $10.7$ dB SI-SDR using the same separator and pipeline.
Replacing IVA with spatial clustering leads to failure ($-0.8$ dB). A possible explanation is that the spatial clustering demixer yields lower-quality virtual microphones, providing insufficient high-SNR cues to stabilize learning in this setup with a limited number of microphones.

\section{Conclusion}
\label{sec:conclusion}

We have proposed VM-UNSSOR, an unsupervised speech separation algorithm that augments the physical array with higher-SNR virtual microphones formed by linear spatial demixers.
As linear projections of the observed mixtures, the virtual microphone signals satisfy mixture consistency and increase source dominance.
By combining physical and virtual microphones and enforcing mixture consistency, VM-UNSSOR enlarges the number of constraints and remains effective in determined conditions.
On SMS-WSJ, VM-UNSSOR clearly outperforms UNSSOR.
In the $6$-microphone, $2$-speaker setup, it reaches $17.1$ dB SI-SDR and $18.0$ dB SDR.
In the determined $2$-microphone $2$-speaker setup, UNSSOR fails to train ($-2.7$ dB SI-SDR), while VM-UNSSOR attains $10.7$ dB.
VM-UNSSOR requires no labeled sources and no additional hardware, making it attractive for rapid in-domain adaptation. 

\newpage

{\footnotesize
\bibliographystyle{IEEEtran}
\bibliography{references}
}

\end{document}